\documentclass[11pt,twoside]{article}
\usepackage{./asp2014}

\aspSuppressVolSlug
\resetcounters

\begin{document}

\title{A new outburst of the yellow hypergiant star $\rho$\,Cas}
\author{Anna Aret,$^1$ Michaela Kraus,$^{1,2}$ Indrek Kolka,$^1$ and Grigoris Maravelias$^2$
\affil{$^1$Tartu Observatory, 61602 T{\~o}ravere, Tartumaa, Estonia; \email{aret@to.ee}\\
$^2$Astronomick\'y \'ustav AV\v{C}R, Fri\v{c}ova 298, 25165 Ond\v{r}ejov, Czech Republic}}

\paperauthor{Anna Aret}{aret@to.ee}{}{Tartu Observatory}{}{T{\~o}ravere}{Tartumaa}{61602}{Estonia}
\paperauthor{Michaela Kraus}{kraus@to.ee}{}{Tartu Observatory}{}{T{\~o}ravere}{Tartumaa}{61602}{Estonia}
\paperauthor{Indrek Kolka}{indrek@to.ee}{}{Tartu Observatory}{}{T{\~o}ravere}{Tartumaa}{61602}{Estonia}
\paperauthor{Grigoris Maravelias}{maravelias@asu.cas.cz}{}{Astronomick\'y \'ustav AV\v{C}R}{}{Ond\v{r}ejov}{}{25165}{Czech Republic}

\begin{abstract}
Spectroscopic monitoring of the yellow hypergiant $\rho$\,Cas revealed a new outburst in 2013, which is obvious from the development
of TiO bands in the spectra. Also many atmospheric lines characteristic for a later spectral type appear. This spectroscopic outburst is in agreement with the photometric light curve, which displays a drop by about 0.6 mag during the same period.
\end{abstract}

\section*{Introduction}\label{introduction}
Yellow hypergiants are massive stars that have passed through the red-supergiant phase. 
On their way back bluewards in the Hertzsprung-Russell diagram (HRD) their envelopes 
become unstable as soon as the effective temperature reaches a value of 7000~K, and the star experiences an outburst.
The yellow hypergiant $\rho$\,Cas is famous for its outbursts, 
during which the star develops TiO bands in a cool, optically thick wind with a very
brief but high mass-loss rate \citep{2003ApJ...583..923L}. 
Each outburst is accompanied by a steep drop in the light curve.
At least three such outbursts were recorded for $\rho$\,Cas: 
1945--1947, 1985--1986, and 2000--2001. 

In 2010 we started to monitor spectroscopically several yellow hypergiants 
using the Ond\v{r}ejov 2m telescope. The aim of this campaign is to track and 
study their mass ejection phases. One of the objects we monitor is 
$\rho$\,Cas. Our spectroscopic data show that during 2013 another outburst occurred. 
The decrease of the time interval between the outbursts might indicate that $\rho$\,Cas is 
preparing for its passage through the Yellow Void region \citep{1997MNRAS.290L..50D} towards the 
hot side of the HRD.

\section*{Observations and results}\label{results}
The observations were obtained during 2010--2015 using the Coud\'{e} spectrograph attached
to the Perek 2-m telescope at Ond\v{r}ejov Observatory \citep{2002PAICz..90....1S}.
Spectra were taken in the wavelength regions
6250--6760\,{\AA} ($R\simeq$ 13\,000) and 
6990--7500\,{\AA} ($R\simeq$ 15\,000). 

\articlefigure{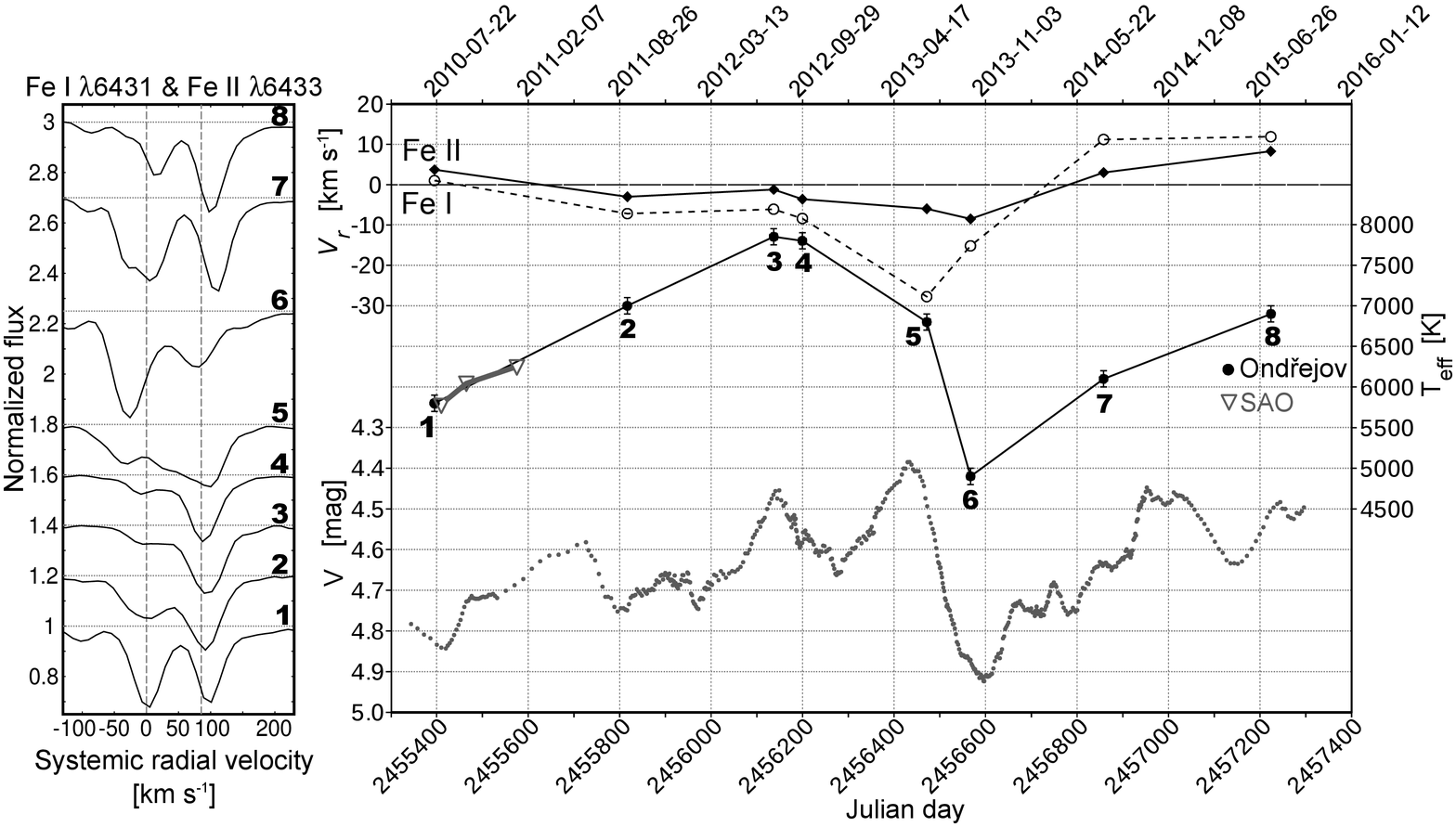}{fig1}{\emph{Left:} Profiles of the Fe\,I and Fe\,II lines in the years 2010--2015. A systemic velocity of --47 km\,s$^{-1}$ was adapted.
 \emph{Right:} V-brightness curve of $\rho$\,Cas (gray dots) is compared with effective temperature $T_{\rm eff}$ values and with radial velocities $V_r$ of the Fe\,{\sc i} $\lambda$6412 and Fe\,{\sc ii} $\lambda$6443 lines.}

The spectra taken on October 1, 2013 reveal weak TiO absorption bands at 7055\,{\AA} and 7126 \,{\AA} 
 together with many atmospheric lines characteristic for a later spectral type.
Photometric light curve obtained from the AAVSO International Database and BAAVSS database (contributed by W.~Vollmann and D.~Loughney, respectively) shows a drop in brightness by about 0.6 mag from May to October 2013.
Changes in the line profiles are remarkable already in the spectra taken in June 2013 (data point 5 in Fig.\ref{fig1}), 
when the visual brightness was still high. Note the broad blue wings of the Fe\,{\sc i} and Fe\,{\sc ii} lines indicating the onset of a strong stellar wind.

We estimated the temperature of $\rho$\,Cas (right panel in Fig.\ref{fig1}) using the Fe\,{\sc i} $\lambda$6431 / Fe\,{\sc ii}  $\lambda$6433 line ratio (left panel in Fig.\ref{fig1}), calibrated by high resolution spectra (Elodie and UVES POP)
of 19 late A to early K supergiants with precise effective temperature values $T_{\rm eff}$ from the literature  \citep{2007MNRAS.378..617K}. The estimated error is $\pm$100\,K.
Variations of $T_{\rm eff}$ follow the general trend of the light curve: hotter during bright phases and cooler during the outburst. Our temperature estimates agree with the results by \citet{2014ARep...58..101K} during the time overlap of our observations in 2010--2011 (SAO spectra: empty triangles in Fig.\ref{fig1}).

The radial velocity variability displayed by lines of different elements in various ionization stages indicates the vertical velocity structure within the outer atmosphere. The amplitude in Fe\,{\sc i} is larger than in Fe\,{\sc ii}: the outer layers where Fe\,{\sc i} lines form are much more affected by the outburst activity. We observe the same shift of the Fe\,{\sc i} radial velocity curve with respect to the light curve as reported by \citet{2003ApJ...583..923L}, while the  Fe\,{\sc ii} radial velocity curve follows both the light and effective temperature curves.

\acknowledgements A.A. and I.K. acknowledge financial support from Estonian grant IUT40-1; M.K. and G.M. from GA\,\v{C}R (14-21373S) and  RVO:67985815.

\end{document}